\documentclass[
	aps, prb,
	superscriptaddress,
	epsf,
	amsmath, amssymbol,
	twocolumn,
	]{revtex4-1}

\usepackage{graphicx}
\def\deg{\ensuremath{^{\circ}}}

\begin{document}


\title{Successive magnetic ordering sequence observed from muon spin rotation in T'-structured $R_2$CuO$_4$ ($R$ = Eu, Nd)}

\author{K. M. Suzuki}
	\thanks{k.m.suzuki@imr.tohoku.ac.jp}
	\affiliation{Institute for Materials Research, Tohoku University, 2-1-1 Katahira, Aoba-ku, Sendai 980-8577, Japan}
\author{S. Asano}
	\affiliation{Department of Physics, Graduate school of Science, Tohoku University, 6-3 Aoba, Aramaki, Aoba-ku, Sendai 980-8578, Japan}
\author{H. Okabe}
	\affiliation{Muon Science Laboratory, Institute of Materials Structure Science, High Energy Accelerator Research Organization (KEK), Tsukuba, Ibaraki 305-0801, Japan}
\author{A. Koda}
	\affiliation{Muon Science Laboratory, Institute of Materials Structure Science, High Energy Accelerator Research Organization (KEK), Tsukuba, Ibaraki 305-0801, Japan}
\author{R. Kadono}
	\affiliation{Muon Science Laboratory, Institute of Materials Structure Science, High Energy Accelerator Research Organization (KEK), Tsukuba, Ibaraki 305-0801, Japan}
\author{I. Watanabe}
	\affiliation{Advanced Meson Science Laboratory, Nishina Center for Accelerator-Based Science, The Institute of Physical and Chemical Research (RIKEN), Wako, Saitama 351-0198, Japan}
\author{M. Fujita}
	\thanks{fujita@imr.tohoku.ac.jp}
	\affiliation{Institute for Materials Research, Tohoku University, 2-1-1 Katahira, Aoba-ku, Sendai 980-8577, Japan}

\date{\today}

\begin{abstract}

The magnetic behavior and effects of oxygen-reduction annealing on the magnetism of T'-structured $R_2$CuO$_4$ ($R$ = Eu and Nd), which is the parent compound of electron-doped cuprate superconductors, have been investigated from muon spin rotation ($\mu$SR) measurements. 
In the muon time spectra of as-sintered $R_2$CuO$_4$, an oscillation component with a small amplitude and a dominant exponential-type decay component were observed for the temperature range between $\sim$150 K and $T_\mathrm{N1}$ (=250--270 K). 
This observation suggests an inhomogeneous magnetic phase of a fluctuating spin state and a locally ordered spin state at these temperatures. 
With decreasing temperature, the small oscillation component disappears, and the exponential one develops below $\sim$150 K. 
Upon further cooling, well-defined oscillation spectra corresponding to uniform long-ranged magnetic order emerge below $T_\mathrm{N2} = 110$ K. 
Except for the absence of the small oscillation component, similar magnetic behavior is confirmed for the annealed compounds. 
Therefore, the successive development of magnetism characterized by the fluctuating spin state over a wide temperature range is a common feature for T'-$R_2$CuO$_4$. 
We discuss a possible picture for the ordering sequence in T'-$R_2$CuO$_4$ and the origin of the small oscillation component in the as-sintered compounds from the viewpoint of local defects. 

\end{abstract}

\vspace*{2em}
\maketitle
\newpage

\section{Introduction}

In the study of the microscopic mechanism of superconductivity induced by carrier doping, elucidation of the ground state of the host materials is a fundamental issue. 
The parent compound of electron-doped copper oxide superconductors $R_{2-x}$Ce$_x$CuO$_4$ ($R$ = Pr, Nd, Sm, Eu) is an antiferromagnetic (AFM) insulator and has been considered to have basically the same magnetic properties as those of hole-doped superconductors La$_{2-x}$Sr$_x$CuO$_4$.\cite{Shirane1987,Endoh1989,Kastner1998}
However, muon spin rotation ($\mu$SR) measurements clarified a distinct magnetic ordering sequence between La$_2$CuO$_4$ with a Nd$_2$CuO$_4$-type structure (T'-structure),\cite{Hord2010} which is characterized by the square planar coordination of Cu$^{2+}$ ions, and La$_2$CuO$_4$ with a K$_2$NiF$_4$-type structure (T-structure) containing CuO$_6$ octahedrons.
In T'-La$_2$CuO$_4$, the relaxation of muon spins starts to develop below $T_\mathrm{N1} \approx 200$ K owing to the gradual slowdown of the spin fluctuations.
Upon further cooling, an oscillation component corresponding to the existence of long-ranged magnetic order emerges at $T_\mathrm{N2} = 115$ K.
This successive ordering sequence is in contrast to the magnetic phase transition of T-La$_2$CuO$_4$ at ~300 K, where $T_\mathrm{N2}$ is coincident with $T_\mathrm{N1}$.\cite{Uemura1987} 
This indicates that there are structural effects on the development of the magnetic order in the parent compounds. 
Although the influences of the crystal structure on the physical properties should be excluded for discussing the genuine electron--hole symmetry for the doped Mott insulator, a simple phase diagram joining electron-doped T'-$R_{2-x}$Ce$_x$CuO$_4$ and hole-doped T-La$_{2-x}$Sr$_x$CuO$_4$ has been so far simply compared.

Furthermore, the emergence of superconductivity in T'-$R_2$CuO$_4$ without Ce substitution was recently reported for thin films and low-temperature-synthesized powder samples.\cite{Tsukada2005,Naito2016,Asai2011,Takamatsu2012}
Nuclear magnetic resonance and $\mu$SR measurements showed dynamic spin correlations in superconducting $R_2$CuO$_4$. \cite{Kojima2014,Adachi2016,Fukazawa2017}
These experimental results suggest that the parent compound with the T'-structure is not an AFM Mott insulator but a metal. 
First-principles calculations indeed showed that a Slater insulator, for which the insulating behavior is governed by long-ranged magnetic order, is a possible ground state of T'-$R_2$CuO$_4$.\cite{Weber2010a} 
Moreover, a mean field calculation of the electronic structure demonstrated the metallic character of T'-$R_2$CuO$_4$, while the T-structured system is a Mott insulator.\cite{Das2009a}
The reason for the distinct ground states of the thin films and high-temperature-synthesized standard samples is understood in terms of the residual oxygen at the apical positions above/below the Cu$^{2+}$ ions after a post-annealing treatment; the excess oxygen near the surface of the thin film can be effectively and moderately removed, while homogeneous and complete removal of the apical oxygens from an entire sample is quite difficult for bulk compounds and the remaining chemical disorder suppresses superconductivity.\cite{Naito2016,Adachi2013}

From this point of view, it is discussed for T'-La$_2$CuO$_4$ (Ref. \onlinecite{Hord2010}) that the magnetic behavior shows evidence for the genuine ground state of ideal T'-$R_2$CuO$_4$ without disorder; that is, the partially remaining chemical disorder induces the magnetic order with a somewhat lower $T_\mathrm{N2}$, and the ordered phase will vanish when the disorder is completely removed.\cite{Adachi2016}
However, since the availability of superconducting T'-$R_2$CuO$_4$ is quite limited and superconductivity has not yet been realized in a bulk single crystal, the ground state of parent materials with the T'-structure is still under debate. 
In this situation, a comprehensive study of even nonsuperconducting bulk samples of parent T'-$R_2$CuO$_4$ is important to gain insight into its characteristic physical properties.
Hence, we performed $\mu$SR measurements, which give information of the local magnetism, of both as-sintered and annealed T'-$R_2$CuO$_4$ to clarify the universality of the magnetic behavior in T'-La$_2$CuO$_4$ and the effect of annealing on the magnetism.

For this purpose, we selected Eu$_2$CuO$_4$ and Nd$_2$CuO$_4$ since the inherent nature of the magnetism of the CuO$_2$ plane could be extracted from experimental results irrespective of the rare-earth ions in the block layer.
The pioneering $\mu$SR measurements by Luke {\it et al}., as well as the following study by Baabe {\it et al}., of Nd$_2$CuO$_4$ examined the overall temperature dependence of the magnetism and determined the ordering temperature corresponding to $T_\mathrm{N1}$.\cite{Luke1990,Baabe2004}
In this paper, we report precise studies on the thermal evolution of magnetism, focusing on the characteristic temperatures $T_\mathrm{N1}$ and $T_\mathrm{N2}$ and the effects of reduction annealing on the magnetic behavior in the parent compounds, which have not been reported in earlier studies.

The organizations of this paper is as follows. Sample preparation and experimental details are described in the next section. The results of zero-field (ZF) and longitudinal-field (LF) $\mu$SR measurements are presented in Section III. Finally, we discuss the universal magnetic behavior and the effect of annealing on the magnetism in T'-$R_2$CuO$_4$ in Section IV, elaborating on possible pictures for the ordering sequence and the effect of annealing.

\section{Experimental details}


Polycrystalline samples of Eu$_2$CuO$_4$ and Nd$_2$CuO$_4$ were prepared by a standard solid-state reaction method. 
A prefired powder was pressed into pellets and sintered in air at 1050\deg C.
We refer to these samples as as-sintered (AS) samples/compounds. 
Then, some of the AS samples were annealed in a high-purity argon gas flow at 700\deg C for 18 h.
The obtained samples are referred as annealed (AN) samples/compounds.
From the weight loss of the samples through the reduction annealing process, we estimated the amounts of removed oxygen to be $\delta \approx 0.02$ for both Eu$_2$CuO$_{4+\xi-\delta}$ and Nd$_2$CuO$_{4+\xi-\delta}$, where $\xi$ is an unknown amount of excess oxygen originally contained in the AS samples. 
The phase purity was checked by X-ray powder diffraction measurements of the ground samples, and we confirmed that both the AS and AN samples were single-phase.
The lattice parameters were determined from the diffraction pattern assuming $I4/mmm$ crystal symmetry and are summarized in Table \ref{Table}.
We note that all of our samples show no superconductivity, even after the annealing procedure. 

\begin{table}[b]
	\caption{Lattice constants and characterized magnetic ordering temperatures $T_\mathrm{N1}$ and $T_\mathrm{N2}$ for as-sintered (AS) and annealed (AN) $R_2$CuO$_4$ ($R$ = Eu and Nd).}
	\label{table_latticeparameter}
	\begin{tabular}{llllll}
		\hline
		&&a (\AA)&c (\AA)&$T_\mathrm{N1}$ (K) & $T_\mathrm{N2}$ (K)\\
		\hline
		Eu$_2$CuO$_4$&AS & 3.9049(1) & 11.9121(3) & 265(10) & 110(10)\\
		&AN & 3.9047(1) & 11.9126(3) & 250(10) & 110(10)\\
		Nd$_2$CuO$_4$&AS & 3.9441(1) & 12.1709(4) & 275(5) & 110(5)\\
		&AN & 3.9458(1) & 12.1639(5) & 258(5) & 110(5)\\
		\hline
	\end{tabular}
	\label{Table}
\end{table}

ZF and LF $\mu$SR measurements were carried out at the Materials and Life Science Facility (MLF) at J-PARC, Japan and at the RIKEN-RAL Muon Facility of the Rutherford Appleton Laboratory (RAL) in the UK using single-pulsed positive surface muon beams. 
The samples were attached to a silver plate and cooled down to 10 K using an open-cycle $^4$He-flow cryostat. 
A $\mu$SR time spectrum---namely, the time evolution of the asymmetry of positron events---is given by $A(t) = [F(t) - \alpha B(t)]/[F(t) + \alpha B(t)]$, where $F(t)$ and $B(t)$ are histograms of the positron event of the detectors that are aligned along the muon momentum direction upstream and downstream relative to the sample, respectively. 
$\alpha$ is a calibration factor for the efficiency and the solid angle of the detectors, and we evaluated it from an oscillation spectrum obtained in the presence of a weak transverse magnetic field at around 300 K for each sample.
The obtained $\mu$SR time spectra were analyzed by using the WiMDA program.\cite{Pratt2000}

Following previous work~\cite{Kubo2002,Fujita2003}, we define $T_\mathrm{N1}$ ($T_\mathrm{N2}$) as the temperature where the ZF-$\mu$SR spectra changes from Gaussian-type decay to exponential-type decay (the oscillation component with a sizable amplitude starts to appear) upon cooling. 
 In Table \ref{table_latticeparameter}, we summarize $T_\mathrm{N1}$ and $T_\mathrm{N2}$  in advance.

\section{Results}\label{Results}
\subsection{Zero-field $\mu$SR measurements}

\begin{figure}[t]
	\includegraphics[width=0.8\linewidth]{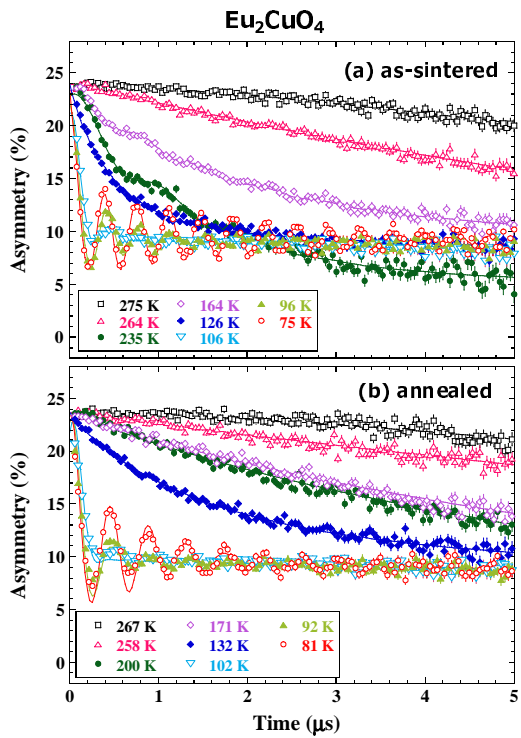}
	\caption{Time spectra of zero-field $\mu$SR for (a) as-sintered and (b) annealed Eu$_2$CuO$_4$. Solid curves are the result of a least-squares fitting by applying Eq. (\ref{eq1})--(\ref{eq3}).}
	\label{ZFspectraECO}
\end{figure}

Figure \ref{ZFspectraECO}(a) shows the time spectra of ZF $\mu$SR for AS Eu$_{2}$CuO$_4$ at representative temperatures. 
A Gaussian-type slow decay is observed at 275 K. 
This is typically observed in a paramagnetic state and is thought to be due to nuclear dipole interactions. 
Upon cooling, an exponential-type component appears, as seen in the spectrum at 264 K, indicating a slowdown of electron spin fluctuations within the $\mu$SR time window of $\sim$$10^{-9}$ s, accompanied by the development of the spin correlation. 
With further cooling, the spectra show faster decay at 235 K and turn into an inverted slower decay at 164 K.
This inverse trend can be interpreted as a reduction in the magnitude of the local magnetic field, an enhancement in the spin fluctuation frequency, or the expansion of the non-/weak magnetic region.
In any case, this is not typical magnetic ordering behavior, where the spin fluctuations slow down monotonically toward the freezing of spins at an ordering temperature.
At temperatures of 164--235 K, furthermore, an oscillation component with a small amplitude is seen on top of the dominant exponential-type decay.
Therefore, a partially ordered spin state coexists in this temperature range with a dynamically fluctuating spin state and/or spin-glass-like state, from which the exponential decay arises.
The small oscillation component disappears in the spectrum at 126 K, resulting in simple exponential-type decay spectra.
Subsequently, an oscillation component with the full volume fraction appears, as seen at 75--106 K. 
The oscillation behavior persists beyond 3 $\mu$s, indicating the formation of coherent long-ranged magnetic order. 
These results suggest that the true magnetic order takes place at a much lower temperature than the previously reported temperature of 265 K (Ref. \onlinecite{Chattopadhyay1994}), and the ordering temperature is rather comparable to $T_\mathrm{N2}$ of 110 K for T'-La$_2$CuO$_4$ (Ref. \onlinecite{Hord2010}).

We next show the ZF $\mu$SR spectra of AN Eu$_2$CuO$_4$ in Fig. \ref{ZFspectraECO}(b). 
Similar to the results of the AS sample, exponential-type depolarization appears at $T_\mathrm{N1} = 250$ K with decreasing temperature. (See the spectra at 200, 171, and 132 K.) 
The spectrum at 171 K shows a slightly but definitely slower time decay than that at 200 K. 
The time spectra show faster decay again at 132 K and finally form into oscillation spectra below $T_\mathrm{N2} = 110$ K.
This magnetic behavior is qualitatively similar to that of the AS sample. 
In addition to the similar ordering sequence, the characteristic properties for the long-ranged ordered phase, such as $T_\mathrm{N2}$, the oscillation frequencies, and the damping rates of oscillation, are almost the same for the two compounds. 
One qualitative difference is the existence/absence of the small oscillation component around 200 K in the AS/AN compounds.
Another difference is the considerably weaker depolarization in the AN compound for $\sim$150 K $<$ T $\leq$ $T_\mathrm{N1}$ compared with that in the AS one. 
These results indicate that the magnetism at higher temperatures is varied considerably by reduction annealing.

For quantitative evaluation of the magnetic behavior, we analyzed the time spectra using the following functions: 
\begin{equation}
\begin{split}
	A(t) = A_0\{
	f_\mathrm{se}e^{-(\lambda _\mathrm{se}t)^\beta}
	+ f_\mathrm{osc} e^{-\lambda _\mathrm{d1}t} \cos (\omega _1t + \phi _1)\\
	+ (1 - f_\mathrm{se} - f_\mathrm{osc}) e^{-\lambda _\mathrm{s}t}\} +A_\mathrm{BG}   \hspace{5mm}
	  (T > T_\mathrm{N2}),
	\label{eq1}
\end{split}
\end{equation}
\begin{equation}
	A(t) = A_0 \{f_\mathrm{osc} G_\mathrm{T}(t) + (1-f_\mathrm{osc})e^{-\lambda _st}\}+ A_\mathrm{BG} 
	  (T \leq T_\mathrm{N2}), 
	\label{eq2}
\end{equation}
where 
\begin{equation}
\begin{split}
	G_\mathrm{T}(t) =
	r_1 e^{-\lambda _\mathrm{d1}t} \cos (\omega _1t + \phi _1)
	+ (1 - r_1) e^{-\lambda _\mathrm{d2}t}. 
	\label{eq3}
\end{split}
\end{equation}
$A_0$ is the initial asymmetry at $t = 0$, and $A_\mathrm{BG}$ is the time-independent background mainly originating from the silver plate.
$A_0$ and $A_\mathrm{BG}$ were fixed throughout the analysis. 
$f_\mathrm{se}$ and $f_\mathrm{osc}$ are the fractions of the stretched exponential component and oscillatory component, respectively.
$\lambda_\mathrm{se}$ and $\beta$ are the relaxation rate and power of the stretched exponential component.
$\lambda _\mathrm{d1}$ and $\lambda _\mathrm{d2}$ are the damping rates of the oscillations, and $r_1$ is the ratio of the damping components.
$\omega _1$ and $\phi _1$ are the frequency of the oscillation and its phase, respectively.

\begin{figure}[t]
	\includegraphics[width=0.75\linewidth]{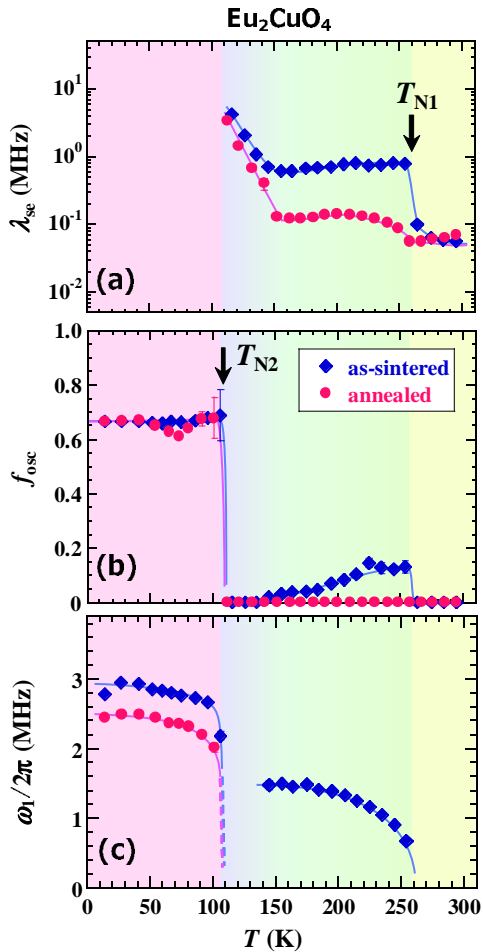}
	\caption{Temperature dependences of (a) the depolarization rate of the stretched exponential term $\lambda _\mathrm{se}$, (b) the asymmetry fraction of oscillation $f_\mathrm{osc}$, and (c) the oscillation frequency $\omega_{1}$ for as-sintered and annealed Eu$_2$CuO$_4$.
	These parameters are obtained by analyzing the spectra using Eq. (\ref{eq1})--(\ref{eq3}).
	All the data for Eu$_2$CuO$_4$ were taken at MLF.
	See the Supplemental Information for the other parameters.}
	\label{paramECO}
\end{figure}

As seen in Fig. \ref{paramECO}(a), the temperature dependences of $\lambda _\mathrm{se}$ for the AS and AN compounds show similar behavior, i.e., a clear jump at $T_\mathrm{N1}$ followed by a hump around 200 K and another increase below $\sim$150 K with cooling.
However, the effects of annealing are seen in the following way; the development of magnetism around $T_\mathrm{N1}$ is broader, and the magnitude of $\lambda _\mathrm{se}$ below $T_\mathrm{N1}$ is much smaller for the AN compound.
For the AS compound, the disappearance of the small oscillation component at 110--150 K is seen as a consequence of the reduction in $f_\mathrm{osc}$ (see Fig. \ref{paramECO}(b)). 
The small oscillation component disappears at $\sim$150 K with the rapid development of $\lambda _\mathrm{se}$ toward $T_\mathrm{N2}$, where the coherent long-ranged magnetic order takes place, suggesting an interrelation between the two components. 
Below $T_\mathrm{N2}$, $\omega_1$ is reduced by a factor of $\sim$0.8 through the annealing. 
Although $\omega_1$ for 110 K $\lesssim T \lesssim$ 150 K is lacking owing to the disappearance of oscillation, the extrapolation of $\omega_1$ at $150\:\mathrm{K} < T < T_\mathrm{N1}$ toward lower temperatures seems not to connect smoothly but rather discontinuous to $\omega_1$ below $T_\mathrm{N2}$.
This discontinuity in $\omega_1$ and the disappearance of the oscillation component just above $T_\mathrm{N2}$ suggest that the oscillation behavior at the high temperatures is not a direct precursor of the static magnetic order below $T_\mathrm{N2}$.

\begin{figure}[t]
	\includegraphics[width=0.8\linewidth]{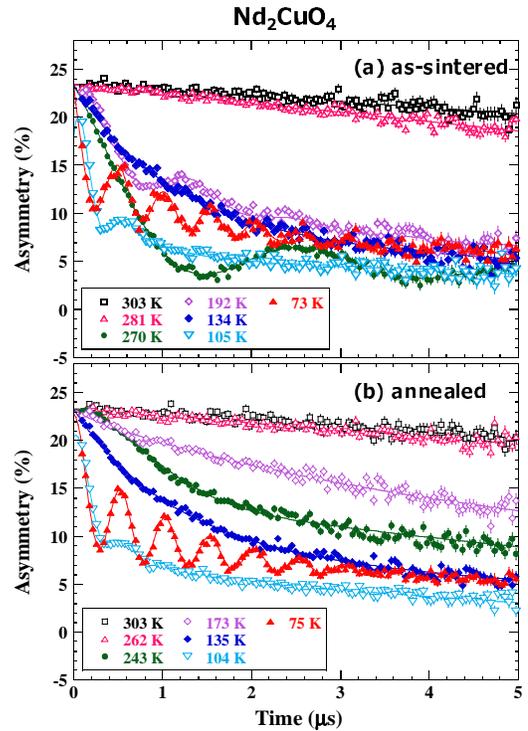}
	\caption{Time spectra of zero-field $\mu$SR for (a) as-sintered and (b) annealed Nd$_{2}$CuO$_4$. Solid curves are the result of a least-squares fitting by applying Eq. (\ref{eq1}), (\ref{eq2}), and (\ref{eq4}). 
	}
	\label{ZFspectraNCO}
\end{figure}

Qualitatively the same results are obtained for Nd$_{2}$CuO$_4$, as shown in Fig. \ref{ZFspectraNCO}(a). 
For the AS compound, the Gaussian-type time spectra showing slow decay above 281 K vary to the exponential-like one at 270 K. 
Thus, $T_\mathrm{N1}$ is determined to be 275 K.
A small but clear oscillation component is observed in the spectra at 270 and 192 K.
The oscillation frequency is larger at 192 K than that at 270 K, while the dominant exponential-type decay component is slower at the lower temperature.
This small oscillation component becomes negligible or vanishes at 134 K, and an oscillation component appears again below $T_\mathrm{N2} = 110$ K (see the spectra at 73 and 105 K in Fig. 3(a)).
The small oscillation component existing above 150 K in AS Nd$_{2}$CuO$_4$ completely disappears by the reduction annealing, while the successive ordering behavior remains in the AN compound, as seen in Fig. \ref{ZFspectraNCO}(b).

We note that there are some differences in the time spectra for AS Nd$_{2}$CuO$_4$ and AS Eu$_2$CuO$_4$.
For example, the time spectra show more rapid depolarization of muon spins around 200 K, and the small oscillation component is more clearly seen in Nd$_{2}$CuO$_4$.
Moreover, in the time spectra of the long-ranged ordered phase below $T_\mathrm{N2}$, the center asymmetry of the oscillation is time-independent for Eu$_2$CuO$_4$, while for Nd$_2$CuO$_4$, it shows a gradual decrease as time evolves.
These differences likely originate from the absence/presence of a rare-earth magnetic moment.
In a wide temperature range, the spectra of Nd$_2$CuO$_4$ contain a dynamic relaxation component even below $T_\mathrm{N2}$, indicating the presence of fluctuating Nd spins.

\begin{figure}[t]
	\includegraphics[width=0.75\linewidth]{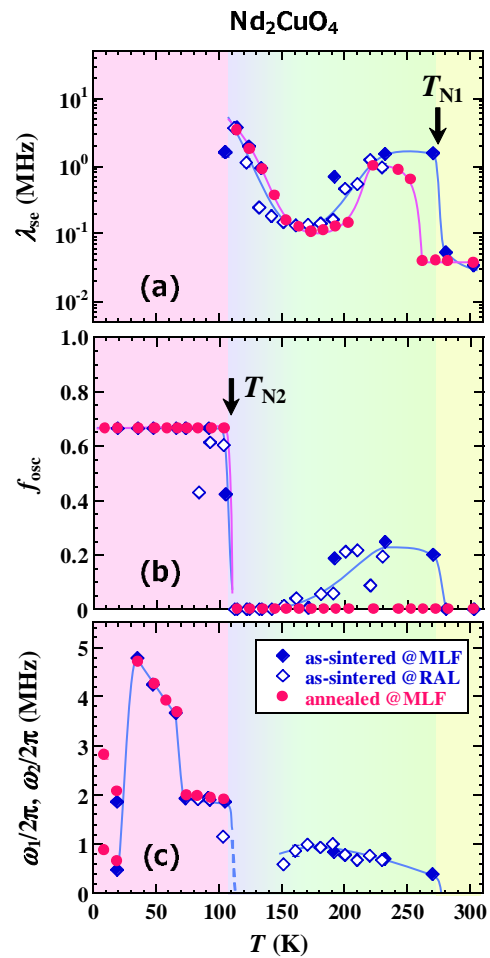}
	\caption{Temperature dependences of (a) the depolarization rate of the stretched exponential term $\lambda _\mathrm{se}$, (b) the asymmetry fraction of oscillation $f_\mathrm{osc}$, and (c) the oscillation frequencies $\omega_{1}$ and $\omega_{2}$ for as-sintered and annealed Nd$_2$CuO$_4$.
	These parameters are obtained by analyzing the spectra using Eq. (\ref{eq1}), (\ref{eq2}), and (\ref{eq4}).
	Open and closed symbols represent the data taken at MLF and RAL, respectively.
	See the Supplemental Information for the other parameters.}
	\label{paramNCO}
\end{figure}

For quantitative analysis, we fitted the spectra by using the same function used for Eu$_2$CuO$_4$ but with
\begin{equation}
\begin{split}
	G_\mathrm{T}(t) =
	r_1 e^{-\lambda _\mathrm{d1}t} \cos (\omega _1t + \phi _1)
	+r_2 e^{-\lambda _\mathrm{d2}t} \cos (\omega _2t + \phi _2)\\
	+(1 - r_1 - r_2) e^{-\lambda _\mathrm{d3}t}
	\label{eq4}
\end{split}
\end{equation}
instead of Eq. (\ref{eq3}). 
$f_\mathrm{osc}$ for AN Nd$_2$CuO$_4$ below $T_\mathrm{N2}$ was fixed to be 2/3, which corresponds to the full volume fraction of oscillation component. 
The results of the optimized parameters are plotted in Fig. \ref{paramNCO}. 
The temperature dependences of the parameters for Nd$_2$CuO$_4$ are qualitatively the same as those for Eu$_2$CuO$_4$.
A prominent difference is seen as the multistep variation in $\omega_1$ at 30 and 75 K, which correspond to the transitions of the spin structure. \cite{Luke1990,Baabe2004}

One of the important findings of this study is that all values of $T_\mathrm{N2}$ are coincident for AS and AN $R_2$CuO$_4$ ($R$ = Eu and Nd), and these values are also comparable to that for T'-La$_2$CuO$_4$. \cite{Hord2010}
Therefore, $T_\mathrm{N2}$ would be independent of oxygen reduction, the lattice constants, and the magnetic moment of the rare-earth ion, and the static order below $T_\mathrm{N2}$ is a robust feature of T'-$R_2$CuO$_4$. 
In contrast, $T_\mathrm{N1}$ varies from 250 K to 280 K depending on $R$ and the annealing conditions. 
The lowest $T_\mathrm{N1}$ of $\sim$200 K was reported for T'-La$_2$CuO$_4$.

\begin{figure*}[t]
	\includegraphics[width=0.7\linewidth]{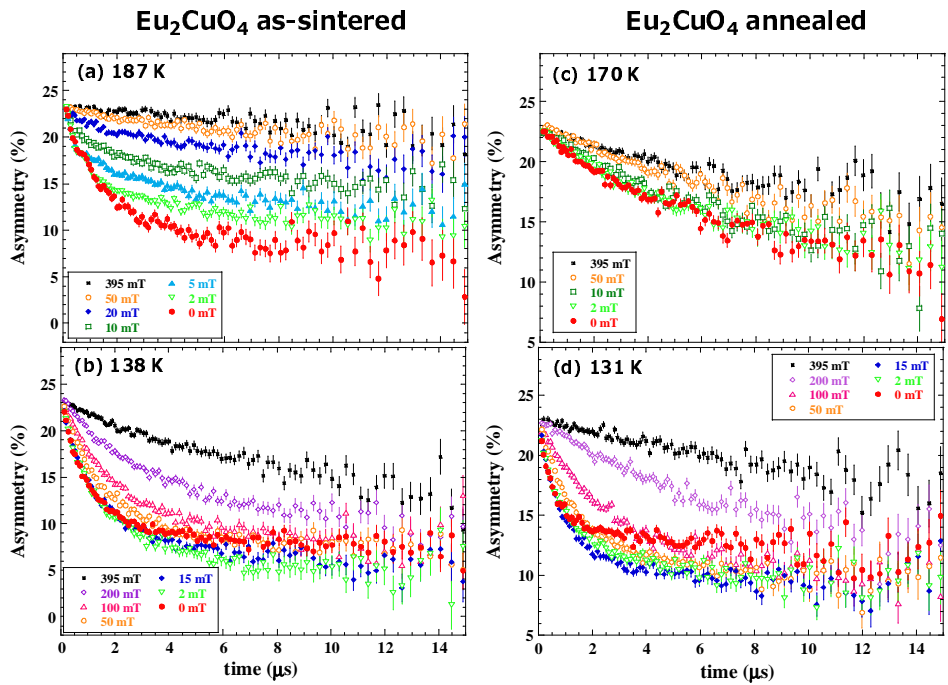}
	\caption{Field dependences of the time spectra of $\mu$SR asymmetries in longitudinal magnetic fields for as-sintered Eu$_2$CuO$_4$ at (a) 187 and (b) 138 K and for annealed Eu$_2$CuO$_4$ at (c) 170 and (d) 131 K. Upward shifts in the asymmetry caused by applying magnetic fields higher than 100 mT are corrected using data taken at $\sim$300 K.}
	\label{LFspectraECO}
\end{figure*}

\subsection{Longitudinal-field $\mu$SR measurements}
LF-$\mu$SR measurements were performed to gain further information on the spin states above $T_\mathrm{N2}$. 
Figures \ref{LFspectraECO}(a) and (b) show the field dependences of the $\mu$SR spectra for AS Eu$_2$CuO$_4$ at 187 and 138 K, respectively.
At 187 K, the spectra in the longer time region exhibit a parallel upward shift by applying magnetic fields.
This shift can be seen in the process of decoupling the interaction between the muon spins and the local spontaneous magnetic fields at the muon sites and gives information on the magnitude of the spontaneous magnetic field.
The resultant magnetic field of the spontaneous and applied longitudinal magnetic fields in a sample eventually becomes parallel to the initial muon spin direction when the applied field is sufficiently increased.
The muon spin polarization hardly relaxes in this strong longitudinal magnetic field limit, giving a time-independent spectrum.

In AS Eu$_2$CuO$_4$ at 187 K, an external field of 20 mT is sufficient to shift the overall spectra over the entire time range, suggesting that the static magnetic fields at the muon sites are less than 20 mT.
This behavior demonstrates the contribution of the static spin state to the dominant exponential decay component at 187 K, indicating a highly damped oscillation component in the spectra.
This type of spectra is often observed in a spin glass state or a static magnetic order with disorder.
However, the slow relaxation that remains even at 395 mT also suggests the existence of a dynamically fluctuating spin state.
Therefore, considering the existence of a small oscillation component at these temperatures, a short-ranged ordered region surrounded by the spin-glass-like spin state would be realized around cores such as defects, and a fluctuating spin state possibly exists outside of them (see Fig. \ref{schematicview}(a)).
As for LF data at 138 K shown in Fig. \ref{LFspectraECO}(b), the spectra are weakly changed with 50 mT, and considerable relaxation remains even at 395 mT.
These results definitely show, surprisingly, that the dynamic nature of the local magnetic field develops at the lower temperature of 138 K. 
This thermodynamically anomalous result will be discussed considering the inhomogeneous spin state in Section \ref{Discussion1}.

The field dependence of the spectra for the AN compound shows similar trends to those observed for the AS compound.
At 170 K, as shown in Fig. \ref{LFspectraECO}(c), the asymmetry is shifted by $\sim$5\% in the entire time range by applying a magnetic field of 50 mT, but it hardly varies with further application of the field.
This field dependence is typical for the inhomogeneous spin state, in which static and dynamic regimes coexist, and is qualitatively the same as that for the AS compound at 187 K. 
The weaker field effect in the AS compound suggests shrinkage of the static magnetic region by the annealing.
This is consistent with the disappearance of the small oscillation by the reduction annealing in the ZF spectra.
At 131 K, as shown in Fig. \ref{LFspectraECO}(d), the field dependence is almost the same as that of the AS data at 138 K, showing a negligible effect of annealing at this temperature.

We note that for the AS (AN) compound at 138 K (131 K), weak magnetic fields of 2--15 mT enhance the relaxation, unlike the typical behavior showing upward shifts in the presence of longitudinal magnetic fields. 
There are a few examples where muon spin relaxation is enhanced by applying longitudinal magnetic fields in frustrated magnets.\cite{Tanaka1996,Higemoto1998a}
It is likely that the weak magnetic field affects and modifies the magnetic state, although what exactly occurs and why this phenomenon occurs at the particular temperature are still under investigation.
At least, dynamic fluctuation is realized at fields over 50 mT.

LF measurements at 80 K show a parallel shift in the asymmetry with no or a negligible dynamic component, indicating the appearance of a static spin state with the full volume fraction at a low temperature, as shown in Fig. S5 in the Supplemental Information.
An analysis of the field dependence of the asymmetry presuming a delta function for the magnetic field distribution yields 16.0 mT as the spontaneous magnetic field, which is in good agreement with 20.4 mT evaluated from the frequency of the oscillation at 80 K.

\section{Discussion}\label{Discussion}

The present results clarify the common features of the temperature dependence of the $\mu$SR time spectra. 
First, the electronic magnetic correlation develops at $T_\mathrm{N1}$, but an exponential-type relaxation appears in the wide temperature range between $T_\mathrm{N1}$ and $T_\mathrm{N2}$. 
A sizable oscillation characteristic to static magnetic order appears only below $T_\mathrm{N2}$.
Second, a small oscillation component exists in the AS compounds below $T_\mathrm{N1}$, and it vanishes just above $T_\mathrm{N2}$ with decreasing temperature. 
This small oscillation component disappears through the oxygen-reduction annealing.

In the next two subsections, we argue a possible origin of the unusual behavior based upon a simple picture, assuming that the time spectra are fully and only attributed to the magnetism of the system. 
The interpretations of the results mentioned in Section \ref{Results} are based on this picture. 
An alternative picture for understanding the thermal evolution of the spectra will be mentioned in Section \ref{Discussion3}.

\subsection{Fluctuating spin state for $T_\mathrm{N2} < T \leq T_\mathrm{N1}$ and static spin state below $T_\mathrm{N2}$}\label{Discussion1}

For simplicity, we first discuss the ordering sequence in the AN compounds, for which the small oscillation component observed for the AS compounds below $T_\mathrm{N1}$ is absent. 
The ZF and LF $\mu$SR measurements of AN $R_2$CuO$_4$ demonstrate that a uniform static magnetic order exists only below $T_\mathrm{N2}$ and that a dynamical fluctuating spin state coexists with a spin-glass-like static state at $T_\mathrm{N2} < T < T_\mathrm{N1}$.
These results are consistent with preceding $\mu$SR studies for T'-$R_2$CuO$_4$.\cite{Luke1990, Baabe2004, Hord2010}
On the other hand, previous neutron diffraction measurements of Eu$_2$CuO$_4$ and Nd$_2$CuO$_4$ showed the appearance of well-defined magnetic Bragg peaks below the N\'eel temperature, which is comparable to $T_\mathrm{N1}$ of the $\mu$SR results, without any indication of an anomaly around $T_\mathrm{N2}$.\cite{Akimitsu1989,Endoh1989,Chattopadhyay1994}
The different temperatures at which the static magnetic order appears, determined by the neutron diffraction and $\mu$SR measurements, likely originate from the different time windows of the probes.
The preceding studies using thermal neutrons\cite{Akimitsu1989,Endoh1989,Chattopadhyay1994} measured magnetic signals integrating an energy range from $-\mathrm{k_B}T$ to $+E_\mathrm{i}$ (incident neutron energy) of $\sim$30 meV corresponding to $\lesssim$10$^{-14}$ s in the time window, while $\mu$SR detects slower fluctuations of $\lesssim$10$^{-9}$ s.
Therefore, the fluctuating spin state with a time scale of $\sim$10$^{-9}$ to $\sim$10$^{-14}$ s would be realized for $T_\mathrm{N2} < T \leq T_\mathrm{N1}$.

The magnetic interaction in the $R_2$CuO$_4$ system is basically understood as a two-dimensional Heisenberg magnet with weak XY anisotropy. 
Neutron diffraction measurements suggest that the three-dimensional ordering behavior at $T_\mathrm{N1}$ is governed by the XY anisotropy due to the fully frustrated interplane Cu-spin correlation in the tetragonal crystal symmetry.\cite{Matsuda1990, Keimer1992a}
Thus, one of the possible fluctuating states is that the moment directions are confined in the CuO$_2$ plane but slightly fluctuate with the AFM correlations within the plane around particular axes, which is identical to the moment directions in the ordered phase below $T_\mathrm{N2}$. 
$T_\mathrm{N1}$ would correspond to the temperature at which the confinement of the moment directions takes place upon cooling. 
True static magnetic order is formed below $T_\mathrm{N2}$.

Hord and coworkers reported a reduced $T_\mathrm{N2}$ for T'-La$_2$CuO$_4$ compared to that of T-La$_2$CuO$_4$, for which $T_\mathrm{N2} \approx T_\mathrm{N1}$ (=325 K). 
Our present study showed the same feature of a lower $T_\mathrm{N2}$ for T'-$R_2$CuO$_4$ ($R$ = Eu and Nd).
Therefore, the difference in the crystal structure is most likely the origin of the variation in $T_\mathrm{N2}$ for the parent compounds.
We newly clarified that the magnetic effect from the rare-earth moment on $T_\mathrm{N2}$ for T'-$R_2$CuO$_4$ is negligible.
These new results strongly support the discussion by Hord {\it et al}.; the apical oxygens of the CuO$_6$ octahedra could stabilize the static order of copper spins compared to the planar coordination, resulting in a higher (lower) $T_\mathrm{N2}$ for T-La$_2$CuO$_4$ (T'-$R_2$CuO$_4$).

We should note that we reported $T_\mathrm{N1} = 130$ K for AN Nd$_2$CuO$_4$ (Ref. \onlinecite{Kubo2002}) from early $\mu$SR experiments. 
This lower $T_\mathrm{N1}$ than the present result arises from insufficient temperature steps and rather the poor statistics of the data, which were dominated by the signal of the Nd magnetic moment. 
In previous work, since we confirmed the appearance of an exponential-type decay component below $\sim$150 K, we focused on temperatures below $\sim$200 K considering that $T_\mathrm{N1}$ was 150 K. 
Here, we emphasize that the fine temperature steps over a wide temperature range and the high-quality statistics of the present experimental results successfully clarified the detailed thermal evolution of the Cu moment in Nd$_2$CuO$_4$. 
The existence of a small oscillation component was also revealed by the high-quality data.

\subsection{Locally ordered spin state below $T_\mathrm{N1}$}\label{Discussion2}

We next discuss the locally ordered spin state, which was observed as the presence of a small oscillation component in the AS compounds for $\sim$150 K $< T \leq T_\mathrm{N1}$. 
The most prominent effect of reduction annealing on the magnetism is the disappearance of this partially ordered spin state.

Our recent X-ray absorption near-edge structure (XANES) measurements of T'-Pr$_2$CuO$_4$ revealed an aspect of electron doping in the reduction annealing treatment.\cite{Asano2018}
According to the quantitative analyses of the XANES data, the removal of oxygen of $\delta$ = 0.02 introduces $\sim$0.04 electrons per Cu atom. 
Therefore, one may think that the variation in the magnetism, especially the disappearance of the locally ordered spin state, below $T_\mathrm{N1}$ by annealing originates from the carrier doping. 
However, our preliminary $\mu$SR measurements of AS Eu$_{2-x}$Ce$_x$CuO$_4$ have revealed that the small oscillation component below $T_\mathrm{N1}$ remains at least up to $x = 0.10$ (Ref. \onlinecite{ECCO}).
Therefore, the existence of the small oscillation---namely short-ranged magnetic order---is not related to the doping level but the intrinsic character for compounds that do not experience the reduction annealing. 
The disappearance of this component suggests another aspect of the effect of annealing on the magnetism in addition to the carrier doping.

\begin{figure}[t]
	\includegraphics[width=50mm]{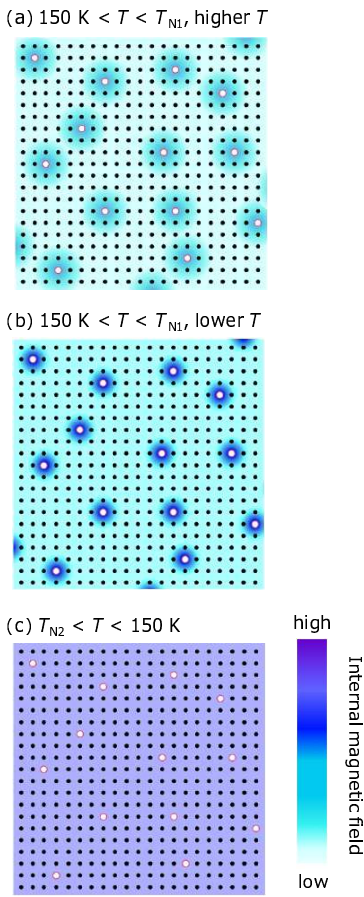}
	\caption{Schematics of the putative spin states on the CuO$_2$ plane for (a) the higher temperature region of $150 \:\mathrm{K} < T < T_\mathrm{N1}$, (b) the lower temperature region of $150 \:\mathrm{K} < T < T_\mathrm{N1}$, and (c) $T_\mathrm{N2} < T < 150$ K.}
	\label{schematicview}
\end{figure}

To understand the microscopic origin of the small oscillation component, the experimental fact that the oscillation appears right below $T_\mathrm{N1}$ is important since it indicates a close connection between the locally ordered region and the dominant spin-glass-like/dynamically fluctuating region. 
As mentioned in Section I, the annealing procedure is recognized to remove the chemical disorder in the AS sample so that an ideal CuO$_2$ plane without a random potential is formed. 
In the slowly fluctuating spin state (for $T_\mathrm{N2} < T \leq T_\mathrm{N1}$ in the present system), chemical disorder---i.e., excess oxygen which partially exists at the apical sites or defects on the CuO$_2$ plane~\cite{Radaelli1994c, Mang2004a, Kimura2005, Kang2007}---could be pinning centers for the fluctuating spins on the underlaying CuO$_2$ plane and induce the local magnetic order. 
This picture is qualitatively consistent with the existence/absence of the oscillation component below $T_\mathrm{N1}$ for the AS/AN samples, and the dynamic nature is enhanced by the annealing. 
For the present AN samples, the nonmonotonic evolution of the muon spin relaxation suggests a tiny amount of disorder remains in the AN samples.

Now, we discuss the disappearance of the small oscillation component and the anomalous enhancement in the dynamic nature of the magnetism of the AS compound for $T_\mathrm{N2} < T \lesssim 150$ K with cooling.
As described in Subsection \ref{Discussion1}, one of the possible fluctuating states is a state with a swing of spins around particular directions within the CuO$_2$ plane.
This state is a precursory state of static magnetic order.
When the correlation length of the fluctuating spin state develops toward the statically ordered state, the partially static state induced by disorder would take part in the fluctuating state since the magnetic energy by the AFM exchange coupling becomes larger than the pinning energy.
Therefore, the volume fraction of the ordered region decreases upon cooling and disappears just above $T_\mathrm{N2}$, where the correlation length of the dominant fluctuating state diverges. 
In this context, the disappearance of the small oscillation component would not occur when the volume fraction of the fluctuating state is markedly suppressed by the large amount of disorder. 
This thermal evolution of the magnetic behavior on the CuO$_2$ plane is schematically drawn in Fig. \ref{schematicview}. 
Alternatively, one may think that the fluctuating state is a state that competes with the static spin state below $T_\mathrm{N2}$. 
In this case, the spin structure of competing state should be different from that of the static spin state; otherwise, no magnetic transition occurs. 
However, no evidence for the reorientation of spins at $T_\mathrm{N2} = 110$ K has been reported for Eu$_2$CuO$_4$ and Nd$_2$CuO$_4$ by neutron scattering measurements.\cite{Matsuda1990, Chattopadhyay1994}

\subsection{Effects of muon diffusion and muon trapping on the spectra}\label{Discussion3}

Since anomalous thermal evolution is present in the $\mu$SR spectra, it is worth considering the effects of muon diffusion and muon trapping by defects, both of which vary with the temperature.
In oxide materials, muons usually stop at particular interstitial positions near oxygen ions, stay at the positions for a while, and hop stochastically to a neighboring stopping site due to thermal vibration.
If the temperature is high and muon hopping occurs frequently, most muons experience changes in the magnetic field within their lifetime, even in a magnetically ordered state, resulting in smearing of the oscillation behavior.
This picture can also explain the present observation above $T_\mathrm{N2}$ and raises the possibility that true static magnetic order is realized below $T_\mathrm{N1}$.
Furthermore, supposing the existence of defects in the sample, muons could travel a longer distance from the initial position and reach defect sites within their lifetime at high temperatures.
If the potential at the defect site is sufficiently deep, muons are trapped there and feel a static magnetic field.
The effect of muon trapping can account for the re-emergence of the oscillation behavior above 150 K in the AS compounds and for the increase in $f_\mathrm{osc}$ upon heating.
The reduction in the number of defects through the annealing procedure results in the disappearance of the small oscillation component.

This alternative picture seems to be able to explain the overall thermal evolution of the time spectra.
However, our preliminary analysis based upon a muon diffusion and trapping model failed to obtain a quantitatively reasonable result for muon hopping rates and the density of trapping sites.
We will proceed with further analyses and will present the details elsewhere.

\section{Summary}

We have performed ZF and LF $\mu$SR measurements for Eu$_2$CuO$_4$ and Nd$_2$CuO$_4$ in order to study the universal features of the magnetic behavior in the parent compound of electron-doped superconductors with the T'-structure. 
The effects of reduction annealing on the magnetism were also studied from the results for AS and AN compounds.

In the ZF $\mu$SR spectra of AS Eu$_2$CuO$_4$, both an exponential decay and an oscillation component with a small amplitude were observed for $\sim$$150 \:\mathrm{K} < T \leq T_\mathrm{N1}$ (=250--280 K), indicating the existence of an inhomogeneous magnetic phase with fluctuating and ordered spin states.
The coinciding appearance of two components suggests a close relation between the dynamic and static spin states.
With decreasing temperature, the small oscillation component vanishes once below $\sim$150 K, and a well-defined oscillation component with the full volume fraction appears below $T_\mathrm{N2} = 110$ K. 
Thus, true static magnetic order is formed below $T_\mathrm{N2}$, while $T_\mathrm{N1}$ is comparable to the ordering temperature reported by neutron scattering measurements.\cite{Matsuda1990, Chattopadhyay1994}
LF $\mu$SR measurements indeed demonstrate the dynamic magnetic nature of the system for $T_\mathrm{N2} \leq T \leq T_\mathrm{N1}$. 
A quite similar development of ZF $\mu$SR spectra was observed over a wide temperature range for AN Eu$_2$CuO$_4$, except for the absence of the small oscillation component. 
This result means that the dominant effect of annealing on the magnetism is the disappearance of the locally ordered spin state.
Furthermore, the weak or negligible effect of annealing on the statically ordered phase below $T_\mathrm{N2}$ suggests the strong stability of the static magnetic order in the parent $R_2$CuO$_4$ against the annealing. 
Qualitatively the same magnetic behavior was confirmed for Nd$_2$CuO$_4$, indicating that the successive development of magnetism is common for T'-$R_2$CuO$_4$ and the local magnetic order inherently existing at high temperatures in the AS compound can be removed by the annealing. 
The ordering sequence in T'-$R_2$CuO$_4$ is different from that in T-La$_2$CuO$_4$, demonstrating the structural effect on the magnetic behavior of the parent compounds. 
This is important for the consideration of the electron--hole asymmetry in the physical properties of doped $R_2$CuO$_4$.

\section*{Acknowledgments}

The $\mu$SR measurements at the Materials and Life Science Experimental Facility of J-PARC were performed under user programs (Proposal Nos. 2013A0084, 2014A0205, 2015MP001, and 2017B0270).
We would like to thank the J-PARC and the RIKEN-RAL staff for their technical support during the experiments.
We are also thankful for helpful discussions with T. Adachi, Y. Ikeda, Y. Koike, K. M. Kojima, Y. Nambu, and K. Yamada.
This work was supported by MEXT KAKENHI, Grant Numbers 15K17689 and 16H02125, and the IMSS Multiprobe Research Grant Program.


%

\end{document}